\documentclass[10pt,onecolumn]{article}
\usepackage{osajnl}
\usepackage{graphicx}
\usepackage{amsmath}

\newcommand{\parti}[2]{\frac{\partial #1}{\partial #2}}

\newcommand{\intall}{\int_{-\infty}^{\infty}}

\begin{document}
\title{Spectral phase conjugation via extended phase matching}

\author{Mankei Tsang}
\date{\today}
\address{Department of Electrical Engineering, 
California Institute of Technology, Pasadena, CA 91125}
\email{mankei@sunoptics.caltech.edu}

\begin{abstract}
It is shown that the copropagating three-wave-mixing parametric
process, with appropriate type-II extended phase matching and pumped
with a short second-harmonic pulse, can perform spectral phase
conjugation and parametric amplification, which shows a threshold
behavior analogous to backward wave oscillation. The process is also
analyzed in the Heisenberg picture, which predicts a spontaneous
parametric down conversion rate in agreement with the experimental
result reported by Kuzucu \textit{et al.}  [Phys. Rev. Lett. {\bf 94},
083601 (2005)]. Applications in optical communications, signal
processing, and quantum information processing can be envisaged.
\end{abstract}

\section{Introduction}
In contrast with the more conventional optical phase conjugation
schemes that perform phase conjugation with spectral inversion
\cite{yariv}, spectral phase conjugation (SPC) is the
phase conjugation of an optical signal in the frequency domain without
spectral inversion. Equivalently, in the time domain, SPC is the phase
conjugation and time reversal of the signal complex pulse envelope
\cite{miller}. SPC is useful for all-order dispersion and nonlinearity
compensation \cite{joubert,tsang2003}, as well as optical signal
processing \cite{marom}. Although SPC has been experimentally
demonstrated using photon echo \cite{echo,echo2}, spectral hole burning
\cite{holeburn,holeburn2}, temporal holography \cite{joubert}, spectral
holography \cite{weiner}, and spectral three-wave mixing (TWM)
\cite{marom_ol}, all the demonstrated schemes suffer from the use of
cryogenic setups, non-realtime operation, or extremely high pump
energy.  Pulsed TWM \cite{tsangOC2004} and four-wave-mixing (FWM)
\cite{miller,tsang2004} processes in the transverse-pumping geometry
have been theoretically proposed to efficiently perform SPC, but
have not yet been experimentally realized. All the holographic and
wave-mixing schemes also have strict requirements on the transverse
beam profile of the signal, limiting their appeal for simultaneous
diffraction and dispersion compensation applications.

There is a correspondence between classical SPC and quantum coincident
frequency entanglement, as shown in Ref.~\citeonline{tsang2005} for
the transversely pumped TWM \cite{walton,tsangOC2004} and FWM
\cite{miller,tsang2004} processes.  It is then interesting to see if
other coincident frequency entanglement schemes are also capable of
performing SPC, when an input signal is present.  This paper studies
one of such schemes, which makes use of extended phase matching
(EPM)\cite{giovannetti} and has been experimentally demonstrated
\cite{kuzucu} in a periodically-poled potassium titanyl phosphate
(PPKTP) crystal \cite{konig}.  It is shown in Section \ref{fourier},
for the first time to the author's knowledge, that this EPM scheme is
indeed capable of performing SPC and optical parametric amplification
(OPA), more efficiently than previous proposals.

The analysis also yields a surprising result, namely that the
parametric gain can be theoretically infinite even for a pump pulse
with finite energy, analogous to backward wave oscillation, where
counterpropagating waves are parametrically coupled and can give rise
to mirrorless optical parametric oscillation
(OPO)\cite{kompfner,heffner,kroll,bobroff,harris,ding,yariv1989}.  The
reason for the similarity is that, in the scheme presented here, even
though the signal and the idler copropagate with the pump pulse in the
laboratory frame, they \emph{counterpropagate in the frame of the
moving pump pulse}, because one is faster than the pump and one is
slower.  Hence the moving pump pulse provides both an effective cavity
and parametric gain, leading to oscillation.  In reality, however, the
interaction among the pulses should be ultimately limited by the
finite device length.  It is shown in Section \ref{laplace}, with a
Laplace analysis, that the parametric gain should abruptly increase
above the threshold, where infinite gain is predicted by the Fourier
analysis, but a finite medium length would always limit the gain to a
finite value. Still, as previous proposals of TWM mirrorless OPO have
never been experimentally achieved due to the requirement of a
continuous-wave (CW) pump and the difficulty in phase matching
counterpropagating waves, the presented analysis suggests the exciting
possibility that mirrorless OPO can be realized with an ultrashort
pump pulse and a practical poling period for phase matching of
copropagating modes, if a long enough medium can be fabricated and
parasitic effects can be controlled.  By analyzing the scheme in the
Heisenberg picture in Section \ref{quantum}, a high spontaneous
parametric down conversion rate is also predicted, in excellent
agreement with the experimental result reported in
Ref.~\citeonline{kuzucu}. The result should be useful for many quantum
information processing applications, such as quantum-enhanced
synchronization \cite{giovannetti_nat} and multiphoton entanglement
for quantum cryptography \cite{durkin}. Finally, numerical results are
presented in Section \ref{numerical}, which confirm the theoretical
predictions.

\section{Setup}

\begin{figure}[htbp]
\centerline{\includegraphics[width=0.48\textwidth]{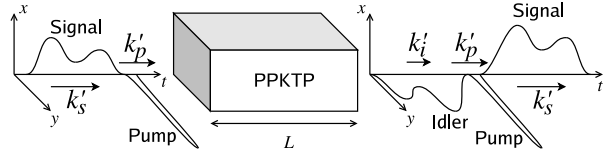}}
\caption{Schematic of spectral phase conjugation (SPC) via type-II
extended phase matching (EPM).  The signal and idler pulses, in
orthogonal polarizations, have carrier frequencies of $\omega_s$ and
$\omega_i$, while the pump pulse has a carrier frequency of $\omega_p
= \omega_s + \omega_i$. The EPM condition requires that the signal and
the idler counterpropagate with respect to the pump, which should be
much shorter than the input signal.}
\label{epm_spc}
\end{figure}

Consider the copropagating TWM process (Fig.~\ref{epm_spc}), assuming
that the basic type-II phase matching condition
($k_s+k_i=k_p+2\pi/\Lambda$), with a quasi-phase-matching period
$\Lambda$, is satisfied. The coupled-mode equations are
\begin{align}
\parti{A_p}{z} + k_p'\parti{A_p}{t} &= j\chi_p A_s A_i, \label{pump0}\\
\parti{A_s}{z} + k_s'\parti{A_s}{t} &= j\chi_s A_p A_i^*,\label{signal0}\\
\parti{A_i^*}{z} + k_i'\parti{A_i^*}{t} &= -j\chi_i A_p^* A_s,\label{idler0}
\end{align}
where $A_p$ is the pump pulse envelope of carrier frequency
$\omega_p$, $A_{s,i}$ are the signal and idler envelopes of frequency
$\omega_s$ and $\omega_i$ respectively, $k'_{p,s,i}$ are the group
delays of the three modes, $\chi_{p,s,i} \equiv \omega_{p,s,i}
\chi^{(2)}/(2cn_{p,s,i})$ are the nonlinear coupling coefficients,
$\omega_{p,s,i}$ are the center frequencies of the modes such that
$\omega_s + \omega_i = \omega_p$, and $n_{p,s,i}$ are the refractive
indices. Group-velocity dispersion within each mode and diffraction
are neglected. Define $\tau \equiv t-k_p'z$ as the retarded time
coordinate that follows the propagating pump pulse.  The change of
coordinates yields
\begin{align}
\parti{A_p}{z} &= j\chi_p A_s A_i,\label{pump}\\
\parti{A_s}{z} + (k_s'-k_p')\parti{A_s}{\tau}
&= j\chi_s A_{p} A_i^*,\label{signal}\\
\parti{A_i^*}{z} + (k_i'-k_p')\parti{A_i^*}{\tau}
&= -j\chi_i A_{p}^* A_s.\label{idler}
\end{align}
Throughout the theoretical analysis, the pump is assumed to be undepleted
and unchirped, so that $A_p = A_{p0}(t-k_p'z) = A_{p0}(\tau)$, hereafter
regarded as real without loss of generality.

\section{\label{fourier}Fourier Analysis}
Equations (\ref{signal}) and (\ref{idler}) are
space-invariant, if the nonlinear medium length $L$ is much longer
than the signal or idler spatial pulse width in the frame of $z$ and
$\tau$, or 
\begin{align}
L >> \frac{T_{s,i}}{|k_{s,i}'-k_p'|},\label{longmedium}
\end{align}
where $T_{s,i}$ is the signal or idler pulse width.  One can then
perform Fourier transform on the equations with respect to $z$, as
defined by the following,
\begin{align}
\tilde{A_s}(\kappa,\tau) &\equiv
\intall A_s(z,\tau)\exp(-j\kappa z) \textrm{d}z,\\
\tilde{A_i^*}(\kappa,\tau) &\equiv
 \intall A_i^*(z,\tau)\exp(-j\kappa z) \textrm{d}z.
\end{align}
Notice that $\tilde{A_i^*}$ is defined as the Fourier transform
after the conjugation of $A_i$. The coupled-mode equations become
\begin{align}
j\kappa \tilde{A_s} + (k_s'-k_p')\parti{\tilde{A_s}}{\tau}
&= j\chi_s A_{p0}(\tau) \tilde{A_i^*},\\
j\kappa \tilde{A_i^*} +(k_i'-k_p')\parti{\tilde{A_i^*}}{\tau}
&= -j\chi_i A_{p0}(\tau) \tilde{A_s}.
\end{align}
Let
\begin{align}
\gamma_s \equiv k_s'-k_p',\mbox{ }
\gamma_i \equiv k_i'-k_p',\mbox{ }
r \equiv \Big|\frac{\gamma_s\chi_i}{\gamma_i\chi_s}\Big|.
\end{align}
Consider the case in which $\gamma_s$ and $\gamma_i$ are non-zero
and have opposite signs, implying that the signal and the idler
propagate in opposite directions with respect to the pump. This can be
achieved for a range of wavelengths in KTP.  Without loss of
generality, assume that $\gamma_s > 0$ and $\gamma_i < 0$, so that
$k_s' > k_p' > k_i'$.  Making the following substitutions,
\begin{align}
A &= \sqrt{r}\tilde{A_s}\exp(j\frac{\kappa}{\gamma_s} \tau),\mbox{ }
B = \tilde{A_i^*}\exp(j\frac{\kappa}{\gamma_i}\tau),
\end{align}
one obtains
\begin{align}
\parti{A}{\tau} &=
j\sqrt{\Big|\frac{\chi_s\chi_i}{\gamma_s \gamma_i}\Big|}A_{p0}(\tau) B
\exp\Big[j\kappa(\frac{1}{\gamma_s}-\frac{1}{\gamma_i})\tau\Big],
\label{A}\\
\parti{B}{\tau} &=
j\sqrt{\Big|\frac{\chi_s\chi_i}{\gamma_s \gamma_i}\Big|} A_{p0}(\tau) A
\exp\Big[-j\kappa(\frac{1}{\gamma_s}-\frac{1}{\gamma_i})\tau\Big].
\label{B}
\end{align}
Due to linear space invariance, the wave-mixing process cannot
generate new spatial frequencies ($\kappa$) for $A$ and $B$.  The
magnitude of $\kappa$ then depends only on the initial bandwidths of
$A$ and $B$, and is on the order of $2\pi\gamma_{s,i}/T_{s,i}$.  As a
result, if the pump pulse width $T_p$ is much shorter than the minimum
period of the detuning factor $\exp[\pm j\kappa
(1/\gamma_s-1/\gamma_i)\tau]$, or
\begin{align}
T_p &<< \Big|\frac{2\pi}{\kappa (1/\gamma_s-1/\gamma_i)}\Big| \sim
\Big|\frac{T_{s,i}}{\gamma_{s,i}(1/\gamma_s-1/\gamma_i)}\Big|,\label{shortpump}
\end{align}
the pump can effectively sample the detuning factor, say, at $\tau =
0$. Defining a normalized coupling function,
\begin{align}
g(\tau) &\equiv
\sqrt{\Big|\frac{\chi_s\chi_i}{\gamma_s \gamma_i}\Big|} A_{p0}(\tau),
\end{align}
two simple coupled-mode equations are obtained,
\begin{align}
\parti{A}{\tau} &= jg(\tau) B, \label{a}\\
\parti{B}{\tau} &= jg(\tau) A. \label{b}
\end{align}
Because the signal and the idler counterpropagate with respect to the pump,
the signal should begin to mix with the pump at the leading edge of
the pump pulse, say at $\tau = -T_p/2$, while the idler should begin
to mix at the trailing edge of the pump, say at $\tau = T_p/2$.
The solutions of Eqs.~(\ref{a}) and (\ref{b}) can then be written as
\begin{align}
A(\kappa,\tau)&=\sec (G)
 \bigg\{A(\kappa,-\frac{T_p}{2})
\cos\Big[\int_{T_p/2}^{\tau} g(\tau') \textrm{d}\tau'\Big]+
jB(\kappa,\frac{T_p}{2})
\sin\Big[\int_{-T_p/2}^{\tau} g(\tau') \textrm{d}\tau'\Big]\bigg\},\\
B(\kappa,\tau) &=
\sec(G) \bigg\{jA(\kappa,-\frac{T_p}{2})
\sin\Big[\int_{T_p/2}^{\tau} g(\tau') \textrm{d}\tau'\Big]+
B(\kappa,\frac{T_p}{2})
\cos\Big[\int_{-T_p/2}^{\tau} g(\tau') \textrm{d}\tau'\Big]\bigg\},
\end{align}
where 
\begin{align}
G \equiv \int_{-T_p/2}^{T_p/2} g(\tau)
d\tau \approx \intall g(\tau) \textrm{d}\tau.
\end{align}
The input signal pulse is required to be placed in advance of
the pump (by $t_s >> T_s$), and the input idler pulse to be placed
behind the pump (delayed by $t_i >> T_i$), so that the signal and the
idler only overlap the pump pulse inside the nonlinear medium.
Consequently, the output solutions are
\begin{align}
A_s(L,t) &= A_{s0}(t-k_s'L+t_s)
\sec(G)+ 
j\frac{1}{\sqrt{r}}A_{i0}^*\big(-\frac{1}{r}(t-k_s'L-t_i)\big)
\tan(G),\label{transforms}\\
A_i(L,t) &=A_{i0}(t-k_i'L-t_i)\sec(G)+
j\sqrt{r}A_{s0}^*\big(-r(t-k_i'L+t_s)\big)\tan(G).\label{transformi}
\end{align}
To see how the device is able to perform SPC, assume that the center
frequencies of the two modes are the same, $\omega_s = \omega_i$,
$\chi_s = \chi_i$, and the type-II EPM condition,
\begin{align}
k_s' + k_i' = 2k_p',\mbox{ }k_s' \neq k_i',
\end{align}
which depends on the material dispersion properties and typically
occurs at a single set of center frequencies, is satisfied
\cite{giovannetti}.  Then $r = 1$, and the output idler becomes the
phase-conjugated and time-reversed replica of the input signal, if the
input idler is zero. SPC is hence performed. The SPC efficiency
$\eta$, or the idler gain, defined as the output idler fluence divided
by the input signal fluence, is
\begin{align}
\eta \equiv \frac{\intall |A_i(L,t)|^2 \textrm{d}t}
{\intall |A_s(0,t)|^2 \textrm{d}t} = \tan^2(G).
\end{align}
This SPC efficiency can be fundamentally higher than that of the
transversely pumped TWM device \cite{tsangOC2004} due to two
reasons. One is the copropagation of the three pulses, which
makes $G$ higher than a similar parameter in the latter case by a
factor of $(1-k_p'/k_s')^{-1}$, on the order of 40 for KTP.  The
second reason is that for $\eta > 1$, due to the tangent function
dependence, the SPC efficiency of the EPM scheme increases with
respect to $G$ much faster than that of the latter, which only depends
on a similar parameter exponentially. That said, the transversely
pumped FWM device \cite{tsang2004} can still be more efficient in the
small gain regime $\eta < 1$ if a highly nonlinear material, such as
polydiacetylene, is used. Furthermore, the EPM device requires a longer
nonlinear medium length by a factor of $(1-k_p'/k_s')^{-1}$, and
depends crucially on the material dispersion, thus severely limiting the
flexibility in the choice of operating wavelengths.

Equations (\ref{transforms}) and (\ref{transformi}) are obtained from
the analysis of the coupled-mode equations (\ref{signal}) and
(\ref{idler}), after Fourier transform with respect to $z$ is
performed. The solutions are therefore formally valid only when the
nonlinear medium length $L$ goes to infinity. In practice, in the
moderate gain regime $\eta \sim O(1)$, the approximation given by
Eq.~(\ref{longmedium}) should be adequate, where the length $L$ can
be, say, ten times larger than the signal spatial pulse width in the
frame of $z$ and $\tau$. Numerical analysis in Section \ref{numerical}
will validate the accuracy of the Fourier solutions.

\section{\label{laplace}Laplace Analysis}

Intriguingly, the Fourier solutions, Eqs.~(\ref{transforms}) and
(\ref{transformi}), have the same form as those of backward wave
oscillation
\cite{kompfner,heffner,kroll,bobroff,harris,ding,yariv1989},
suggesting that the device studied here, with an ultrashort pump pulse
and a practical quasi-phase-matching period ($\Lambda = 46$ $\mu$m as
reported in Ref.~\citeonline{konig}), can also perform mirrorless OPO,
as long as $k_{s,i}'>k_p'>k_{i,s}'$.  However, the prediction of
infinite gain is based on the assumption of infinite medium length and
therefore may not be valid. In this case, Laplace transform should be
used.

For the CW-pumped mirrorless OPO schemes, a Laplace analysis
\cite{fisher} with respect to time shows that beyond threshold, poles
appear on the right-hand plane in the Laplace domain, meaning that the
temporal impulse response increases exponentially with time, leading
to self-oscillation when enough time is elapsed.  The same procedures
of utilizing the two-sided Laplace transform \cite{vanderpol} as in
Ref.~\citeonline{fisher} are followed here in order to be consistent
with the relevant literature, but since the proposed scheme is the
opposite limit of the CW devices, the Laplace transform should be
performed with respect to $z$ instead,
\begin{align}
\bar{A_s}(p,\tau) &\equiv \intall A_s(z,\tau)\exp(-pz) dz,\\
\bar{A_i^*}(p,\tau) &\equiv \intall A_i^*(z,\tau)\exp(-pz) dz.
\end{align}
For simplicity but without affecting the qualitative behavior of the
solutions, it is assumed that the pump pulse is square, there is no
input idler, $\gamma = \gamma_s = -\gamma_i$, and $\chi = \chi_s =
\chi_i$. The output solutions in the Laplace domain are then given by
\begin{align}
\bar{A_s}(p,\frac{T_p}{2}) &= 
\frac{\sqrt{1-P^2}\csc(G\sqrt{1-P^2})}
{P + \sqrt{1-P^2}\cot(G\sqrt{1-P^2})}
\bar{A_s}(p,-\frac{T_p}{2}),
\label{laplaces}\\
\bar{A_i^*}(p,-\frac{T_p}{2})
&= \frac{-j}{P+\sqrt{1-P^2}\cot(G\sqrt{1-P^2})}
\bar{A_s}(p,-\frac{T_p}{2}),
\label{laplacei}\\
P &\equiv \frac{p}{\chi A_{p0}},\mbox{ }
G \equiv \chi A_{p0}(\frac{T_p}{\gamma}).
\end{align}
If we let $p = j\kappa$, the transfer functions in
Eqs.~(\ref{laplaces}) and (\ref{laplacei}) are well-known to be
low-pass filters \cite{pepper}, the bandwidth of which decreases as
$G$ increases. If the spatial bandwidth of the input signal, on the
order of $\gamma/T_s$, is much smaller than the bandwidth of the
low-pass filters, the transfer functions can be regarded as flat-top
functions, and by plugging $P = 0$ in Eqs.~(\ref{laplaces}) and
(\ref{laplacei}), the Fourier solutions in Eqs.~(\ref{transforms}) and
(\ref{transformi}) are recovered. For $G << 1$, the transfer functions
are sinc functions with a bandwidth $\sim \gamma/T_p$, so the Fourier
solutions are valid if $T_p << T_s$, which is essentially the same
assumption used in the Fourier analysis, Eq.~(\ref{shortpump}). As $G$
increases and the filter bandwidth decreases, however, the Fourier
solutions become less and less accurate for a finite-bandwidth input
signal.

The poles of the transfer functions, $p_\infty$, can be obtained by
setting the denominator of Eqs.~(\ref{laplaces}) and (\ref{laplacei})
to zero,
\begin{align}
p_\infty + \sqrt{(\chi A_{p0})^2-p_\infty^2}
\cot\Big[G\sqrt{1-p_\infty^2/(\chi A_{p0})^2}\Big] &= 0.\label{poleseq}
\end{align}

\begin{figure}[htbp]
\centerline{\includegraphics[width=0.48\textwidth]{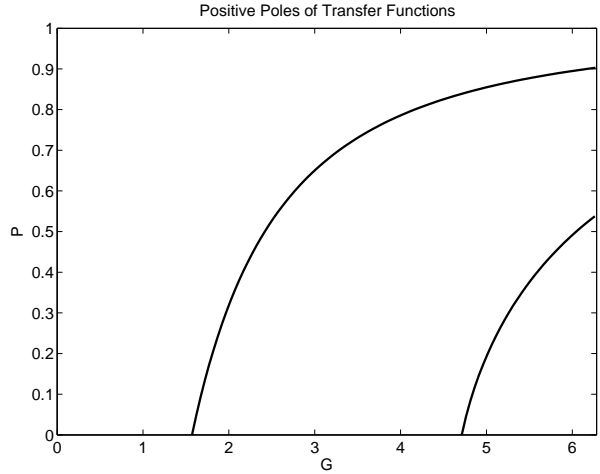}}
\caption{Normalized poles $p_\infty/(\chi A_{p0})$ plotted against $G$,
obtained by numerically solving Eq.~(\ref{poleseq}), indicating the
onset of spatial instability beyond the threshold $G > \pi/2$. More
poles appear as $G$ is increased.}
\label{poles}
\end{figure}

Figure \ref{poles} plots the normalized poles $p_\infty/(\chi A_{p0})$
against $G$. Positive poles begin to appear when $G >\pi/2$, hence the
\emph{spatial} impulse response increases exponentially with respect
to $z$ beyond threshold.

It is interesting to compare the scheme studied here with the case in
which the pump, signal and idler have degenerate group delays
($k_p'=k_s'= k_i'$)\cite{yu}. The coupled-mode equations of the latter
case are
\begin{align}
\parti{A_{s,i}(z,\tau)}{z} &= j\chi A_{p0}(\tau) A_{i,s}^*(z,\tau),
\end{align}
where the $\tau$ derivatives vanish. The solutions are easily
seen to be
\begin{align}
A_{s,i}(z,\tau) &= A_{s,i}(0,\tau)\cosh[\chi A_{p0}(\tau) z] +
jA_{i,s}^*(0,\tau)\sinh[\chi A_{p0}(\tau) z].
\end{align}
This corresponds to the $G\to\infty$ limit of the former scheme, where
$p_\infty/(\chi A_{p0})\to 1$ and all the poles approach the growth
rate of the degenerate case, $\chi A_{p0}$.

\section{\label{quantum}Spontaneous Parametric Down Conversion}
Given the input-output signal-idler relationship in
Eqs.~(\ref{transforms}) and (\ref{transformi}), it is straightforward
to obtain a quantum picture of the parametric process in the moderate
gain regime by replacing the signal and idler envelopes with
Heisenberg operators, so that
\begin{align}
\hat{A}_s &= \hat{A}_{s0}\sec(G)+ j
\hat{A}_{i0}^\dagger\tan(G),\label{operators}\\
\hat{A}_i &=j\hat{A}_{s0}^\dagger\tan(G)+ \hat{A}_{i0}
\sec(G).\label{operatori}
\end{align}
If the inputs are Fock states,
\begin{align}
n_{s,i} &\equiv \langle \hat{A}_{s,i}^\dagger \hat{A}_{s,i}\rangle =
\langle \hat{A}_{s,i} \hat{A}_{s,i}^\dagger\rangle -1,\\
\langle \hat{A}_{s0}^\dagger\hat{A}_{i0}\rangle
&=\langle \hat{A}_{i0}^\dagger\hat{A}_{s0}\rangle
=\langle \hat{A}_{s0}\hat{A}_{i0}^\dagger\rangle
=\langle \hat{A}_{i0}\hat{A}_{s0}^\dagger\rangle = 0.
\end{align}
The average output photon number of each mode is
\begin{align}
n_{s} &= n_{s0}\sec^2(G) + (n_{i0}+1)\tan^2(G),\\
n_{i} &= n_{i0}\sec^2(G) + (n_{s0}+1)\tan^2(G).
\end{align}
The average number of spontaneously generated photon pairs per pump
pulse is therefore the same as the idler gain, or
$\eta=\tan^2(G)$. Moreover, the unitary transform given by
Eqs.~(\ref{operators}) and (\ref{operatori}) has the same form as the
CW FWM process.  One then expects the photon wavefunction to be
similarly given by \cite{fan}
\begin{align}
|\psi\rangle = \cos(G)\sum_{n=0}^\infty\sin^n(G)|n\rangle_s|n\rangle_i,
\end{align}
where $|n\rangle_{s,i}$ is the Fock state in the signal or idler mode.
The scheme thus has a significant advantage in efficiency and
robustness for multiphoton entanglement, compared with other
schemes that often require feedback \cite{lamas}. The efficient 
multiphoton coincident frequency entanglement should be useful for
quantum-enhanced synchronization \cite{giovannetti_nat} and
quantum cryptography applications \cite{durkin}.

The preceding quantum analysis assumes that there is only one spatial
mode in each signal or idler mode, and is accurate only when the
Fourier solutions are accurate. This restricts the applicability of
the quantum analysis to the moderate gain regime $\eta\sim O(1)$,
depending on how closely the assumption in Eq.~(\ref{longmedium}) is
observed. It is beyond the scope of this paper to investigate what
happens in the quantum picture when more than one spatial modes are
involved, but qualitatively, one expects that each spatial mode should
have a varying parametric gain depending on the spatial frequency, as
suggested by the Laplace solutions in Eqs.~(\ref{laplaces}) and
(\ref{laplacei}), so the photon wavefunction would be given by a
superposition of simultaneous eigenstates of spatial frequency and
photon number.

Using the parameters described in Refs.~\citeonline{kuzucu} and
\citeonline{konig}, where $\lambda_0 = 1584$ nm, $\chi^{(2)} = 7.3$
pm/V, $n_0 = 2$, $\gamma = 1.5\times 10^{-10}$ s/m, $T_p = 100$ fs,
average pump power $= 350$ mW, diameter $= 200$ $\mu$m, and pump
repetition rate $f_{rep} = 80$ MHz, the spontaneously generated photon
pairs per second is theoretically given by $f_{rep}\tan^2(G) \approx
f_{rep}G^2 = 3.6\times 10^6$/s, in excellent agreement with the
experimental result reported in Ref.~\citeonline{kuzucu}, which is
$\sim 4\times 10^6$/s.  $G$ is then given by $\sim 0.2$, so the
operations of SPC, OPA, and multiphoton entanglement ($G > \pi/4$)
should be realizable by increasing the pump field amplitude.

\section{\label{numerical}Numerical Analysis}
\begin{figure}[htbp]
\centerline{\includegraphics[width=0.48\textwidth]{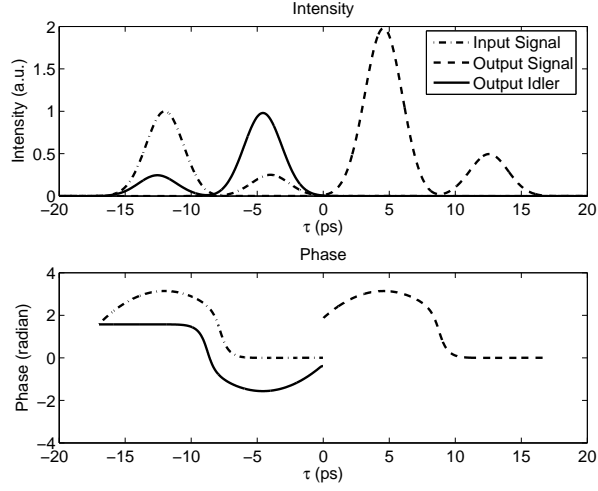}}
\caption{Plots of intensity and phase of input signal, output signal
and output idler, from numerical analysis of Eqs.~(\ref{signal}) and
(\ref{idler}). Parameters used are $k_p'= 1/(1.5\times 10^{8}
\textrm{ms}^{-1})$, $k_s' = 1.025 k_p'$, $k_i = 0.975k_p'$, $T_p =
100$ fs, $T_s = 2$ ps, $L = 10$ cm, $t_s = 4T_s$, beam diameter =
$200$ $\mu$m, $A_{s0} =
0.5\exp[-(t-2T_s)^2/(2T_s^2)]- \exp[-(1+0.5j)(t+2T_s)^2/(2T_s^2)]$,
$A_{p0} = \exp[-t^2/(2T_p^2)]$, and $G = \pi/4$. The plots clearly
show that the idler is the time-reversed and phase-conjugated replica,
i.e. SPC, of the signal.}
\label{intensity_phase}
\end{figure}

Equations (\ref{signal}) and (\ref{idler}) are solved numerically via
a Fourier split-step approach to confirm the above theoretical
predictions. Fig.~\ref{intensity_phase} plots the intensities and
phases of the input signal, output signal, and output idler from the
numerical analysis when $G = \pi/4$. The plots clearly show that the
output idler is the time-reversed and phase-conjugated replica of the
signal. 

\begin{figure}[htbp]
\centerline{\includegraphics[width=0.48\textwidth]{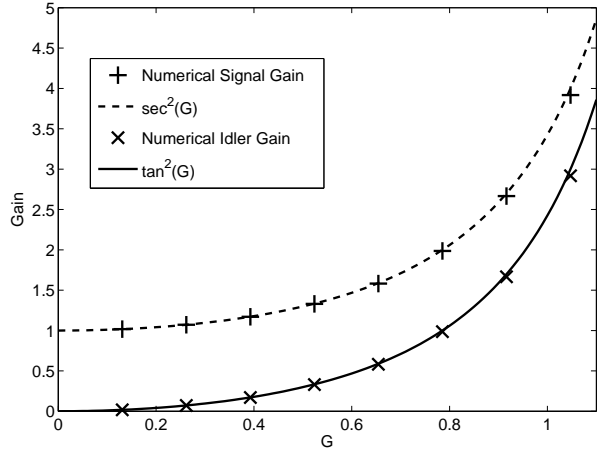}}
\caption{Signal gain $\eta+1$ and idler gain $\eta$ versus $G$
from numerical analysis compared with theory. See caption of
Fig.~\ref{intensity_phase} for parameters used.}
\label{gain}
\end{figure}

Figure \ref{gain} plots the numerical signal gain and idler gain
compared with Fourier theory for $0<G\le\pi/3$. The numerical results are
all within 3\% of the theoretical values.

\begin{figure}[htbp]
\centerline{\includegraphics[width=0.48\textwidth]{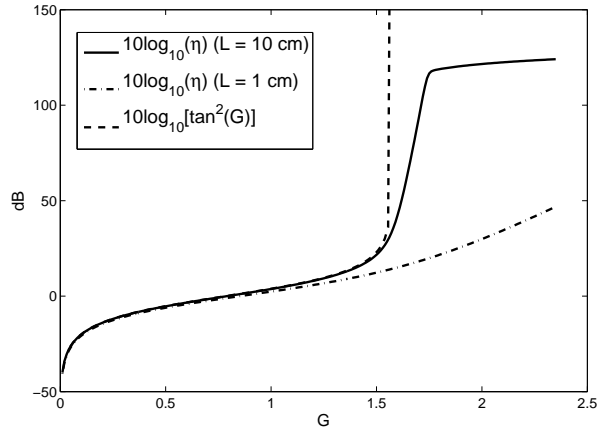}}
\caption{Plot of numerical idler gain $\eta$ in dB against $G$
for $L = 10$ cm (solid) and $L = 1$ cm (dash-dot), compared with the
Fourier theory (dash), $\tan^2(G)$ in dB. Three distinct regimes can
be observed for the $L = 10$ cm case; the moderate gain regime where
the Fourier theory is accurate, the unstable regime where the gain
increases exponentially, and the oscillation regime where significant
pump depletion occurs.  For $L = 1$ cm, the medium is not long enough
for oscillation to occur in the parameter range of interest.}
\label{loggain}
\end{figure}

Figure \ref{loggain} plots the idler gain on the logarithmic scale
for a wider range of $G$'s and two different lengths, obtained from
the numerical analysis of the complete three-wave-mixing equations
(\ref{pump}), (\ref{signal}), and (\ref{idler}), with a single photon
as the input signal, approximately emulating parametric
fluorescence. For the $L = 10$ cm case the curve can be clearly
separated into three regimes; for $G < \pi/2$ and moderate gain ($\eta
\sim 0$ dB), the idler gain approximately follows the Fourier solution
(dashed curve). For $G > \pi/2$, the system becomes unstable and an
exponential growth (linear ramp on the logarithmic curve) is observed,
until the pump is significantly depleted, parametric oscillation
occurs and the exponential growth abruptly stops.

For $L = 1$ cm, the numerical solution departs from theory for a
smaller $G$, and the slope of the logarithmic curve in the unstable
regime, proportional to $L$, is too small to initiate oscillation in
the parameter range of interest.

A medium length of 10 cm may be pushing the limit of current
technology. Even if one is able to fabricate such a long
periodically-poled nonlinear crystal, the effective medium length is
always limited by parasitic effects, such as diffraction,
group-velocity dispersion, and competing third-order nonlinearities,
so it might be difficult to fabricate an ideal EPM device for the
aforementioned purposes. For instance, in the experiment by Kuzucu
\textit{et al.} \cite{kuzucu}, the diameter of the beam is $W \sim
200$ $\mu$m, so the characteristic diffraction length is $\sim
W^2/\lambda_0 = 4$ cm, while the characteristic group-velocity
dispersion length is 20 cm according to Ref.~\citeonline{giovannetti},
which are all on the order of the medium length required for
mirrorless OPO. That said, techniques like diffusion bonding
\cite{michaeli} can be used to increase the length of a nonlinear
crystal, diffraction can be eliminated by waveguiding, while there
exist a variety of methods to compensate for group-velocity dispersion
and third-order nonlinearities \cite{agrawal}. Hence with careful
engineering, fabricating an EPM device for the proposed applications
is still a distinct possibility.

\section{Conclusion}
In summary, it is proven that the copropagating three-wave-mixing
process, with appropriate extended phase matching and pumped with a
short second-harmonic pulse, is capable of performing spectral phase
conjugation, parametric amplification and efficient multiphoton
entanglement. The main technical challenges of experimental
implementation seem to be the long medium length required and the
control of parasitic effects such as diffraction, group-velocity
dispersion, and competing third-order nonlinearities. However, a
shorter proof-of-concept device has already been experimentally
realized for the purposes of broadband second-harmonic generation
\cite{konig} and coincident frequency entanglement \cite{kuzucu}, so
it is not unrealistic to expect that a longer device can be fabricated
for the proposed applications, which should be useful for optical
communications, signal processing, and quantum information processing.

Theoretically, much remains to be explored. The study of parasitic
effects, not considered in this paper, is vital for experimental
realization. The analysis of the ultrashort-pump limit can be
potentially generalized to other TWM and FWM geometries, while the
quantum analysis of this limit is by no means complete. In conclusion,
the analysis presented here should stimulate further experimental and
theoretical investigations of a new class of parametric devices.

The author would like to thank Prof. Demetri Psaltis for helpful
discussions and a reviewer for pointing out
Refs.~\citeonline{kompfner,heffner,kroll,bobroff}.

\end{document}